\documentclass[11pt]{article}
\usepackage{moriond,epsfig}

\def\be{\begin{equation}}
\def\ee{\end{equation}}
\def\bea{\begin{eqnarray}}
\def\eea{\end{eqnarray}}

\begin{document}
\vspace*{4cm}
\title{SIMULATION OF A KM$^3$-SCALE DEEP-SEA NEUTRINO DETECTOR}

\author{ D. ZABOROV \\
(for the ANTARES collaboration)}

\address{Institute for Theoretical and Experimental Physics \\
B. Cheremushkinskaja, 25, 117259 Moscow, Russia}

\maketitle

\abstracts{ 
This document describes a Monte-Carlo simulation of an underwater neutrino 
telescope with a homogeneous detection volume of a cubic kilometre.
}

\section{Introduction}


ANTARES \cite{antprop} is a project of a large area water Cherenkov telescope, 
optimized for measuring the direction of upward-going high-energy muons.
The muons can be induced by high-energy neutrinos of astrophysical origin 
that have traversed the Earth.
The ANTARES detector is currently under construction 
in the deep Mediterranean Sea off shore of Toulon (France).
The telescope will have a sensitive area of the order of 0.1 km$^2$.


 A next step would be the construction of a neutrino telescope 
of effective volume 1 km$^3$ (KM3).
 In order to obtain an estimate of the expected performance of such a detector
we have performed a Monte Carlo simulation study for 
a neutrino telescope containing 8000 optical 
modules on a regular cubic matrix of 20x20x20 with a step size of 60 m. 
Each optical module has a vertically downward-viewing 10-inch photomultiplier 
tube \cite{optmod}.



\section{Simulation}

 The high-energy muons are produced in charged current interactions of
high-energy muon-neutrinos in the surroundings of the detector.
In the simulation the generated muon flux has the following characteristics:

 - Energy distributed as E$^{-1}$ over the energy range 1 TeV - 1000 TeV,

 - Muon tracks are directed upward and distributed isotropically in
    the downward hemisphere.


 The muon interactions with matter as well as the Cherenkov light emission
and propagation have been simulated 
using software developed for the ANTARES project. 
The light absorption, the light scattering and the
background light characteristics were simulated according the parameters
measured at the ANTARES site.\cite{lighttrans,backlight}
The simulated events are passed through the ANTARES reconstruction program.

\vspace{-0.3cm}
\section{Estimate of performance}
\vspace{-0.2cm}

Comparison of the reconstructed tracks with the initial track parameters
allows to determine the main detector characteristics (see Figure 1) as
follows:

 - The effective area of the detector reaches 1 km$^2$ for incoming
muons at 10 TeV energy,

 - The angular resolution is typically 0.07 degrees 
at a muon energy of 10 TeV. \\
 To quantify the angular resolution the median angular difference (error)
between the reconstructed track and the simulated muon track is used.
 A quality cut devised to select optimally well reconstructed events
does neither change the effective area nor the median angular error 
significantly (shaded histogram in Figure 1).

\begin{figure}[h]
\epsfig{file=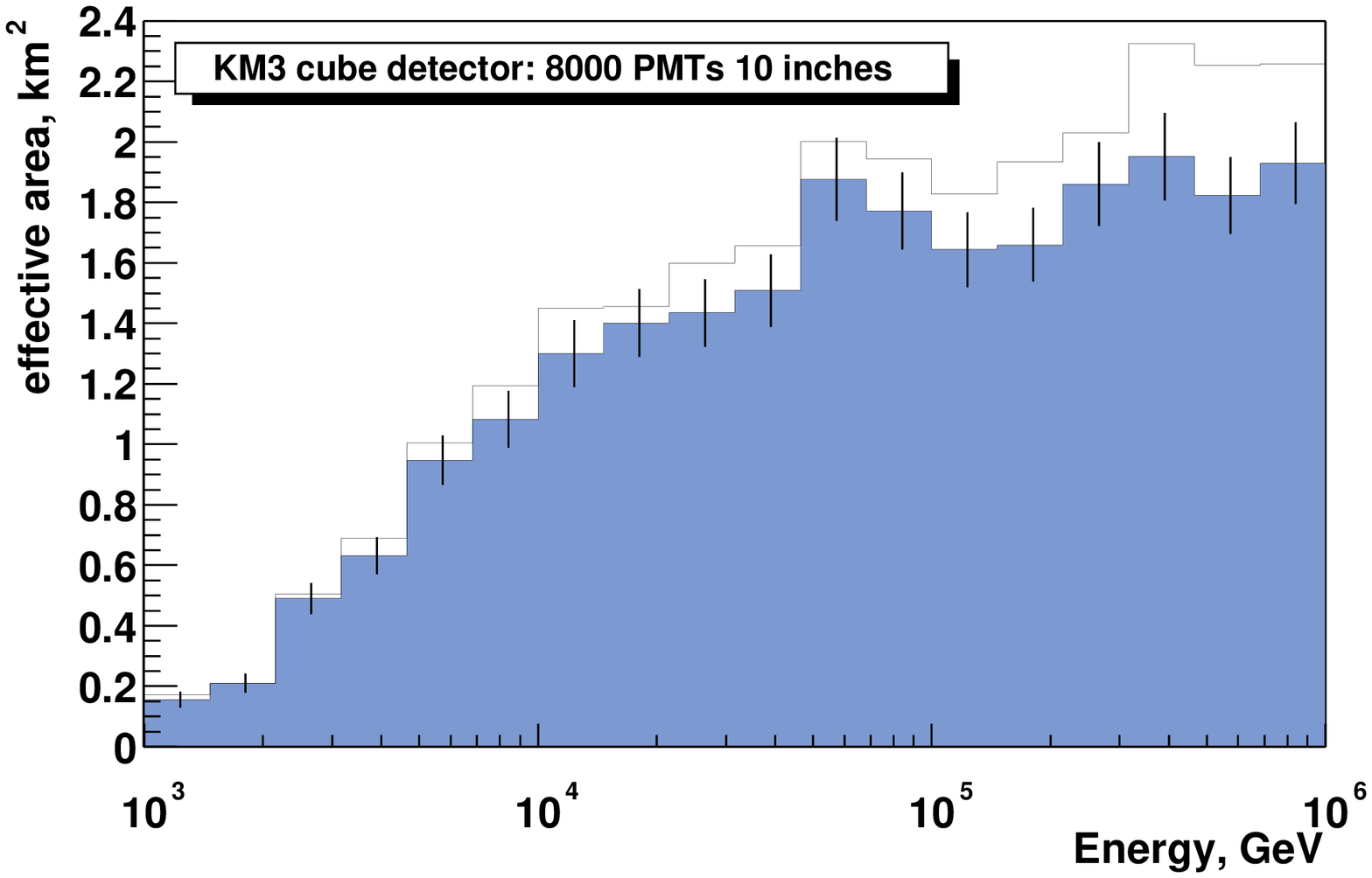,width=7.9cm,height=7.0cm}
\epsfig{file=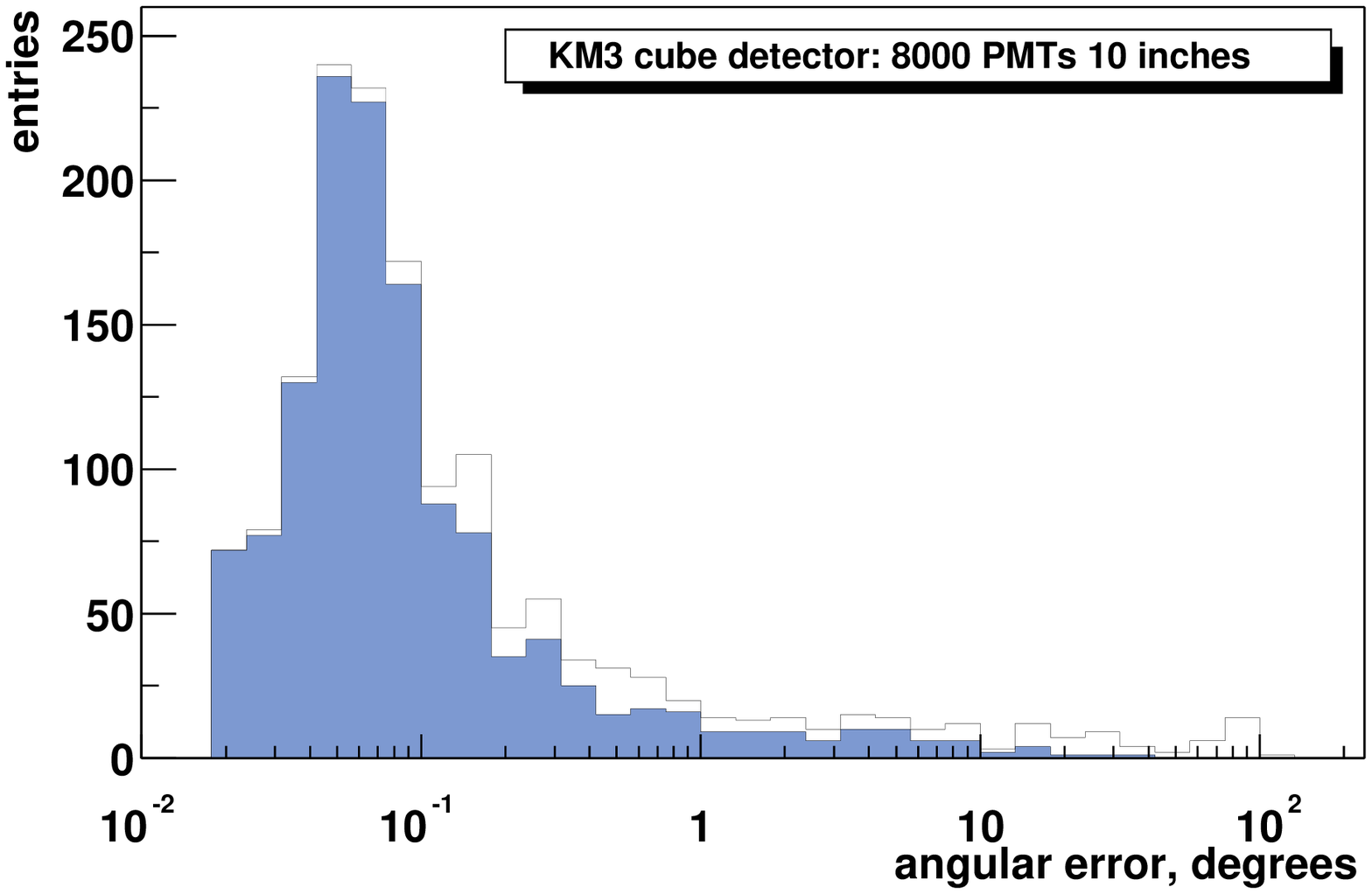,width=7.9cm,height=7.0cm}
\caption{Results of the Monte-Carlo simulation of the km$^3$ detector:
 Effective area of the detector as a function of muon energy (left plot).
 Angular error in muon track reconstruction (right plot).
 The results with the quality cut are shown on the figure as shaded area.
 Error bars on the left plot show statistical errors only.
\label{fig:fig1}}
\end{figure}

\vspace{-0.5cm}
\section{Conclusion}
\vspace{-0.1cm}

 A full Monte-Carlo simulation for a simple model of a neutrino telescope
with a km$^3$ instrumented volume together with muon track reconstruction
shows a good angular accuracy with an effective area that decreases strongly
towards lower muon energies. Optimization of the detector geometry and the
reconstruction algorithms is envisaged as a next step.


\vspace{-0.2cm}
\section*{References}
\begin{thebibliography}{99}
\vspace{-0.2cm}

\bibitem{antprop} 
ANTARES Collaboration, A deep sea telescope for high energy neutrinos, \\
astro-ph/9907432  or http://antares.in2p3.fr.

\bibitem{optmod} P.Amram et al., The ANTARES optical module, 
Nucl. Instr. and Methods A484 (2002) 369 (astro-ph/0112172).

\bibitem{lighttrans} Antares collaboration, Transmission of light in deep 
sea water at the site of the ANTARES neutrino telescope, in preparation.

\bibitem{backlight} P.Amram et al., Background light in potential sites 
for the ANTARES undersea neutrino telescope,
Astroparticle Physics 13 (2000) 127-136 (astro-ph/9910170)


\end{thebibliography}
\end{document}